\documentclass[onecolumn]{svjour2}

\usepackage{natbib}
\usepackage{graphicx}
\usepackage{graphics}

\usepackage{amssymb}

\begin{document}

\title{The Relativistic Factor in the Orbital Dynamics of Point Masses}
\author{Federico Benitez \and Tabar\'e Gallardo}

\institute{Departamento de Astronomia, Instituto de Fisica, Facultad
de Ciencias, Igua 4225, 11400 Montevideo, Uruguay
\email{federico@fisica.edu.uy}}

\maketitle

\begin{abstract}
{There is a growing population of relativistically relevant minor
bodies in the Solar System and a growing population of massive
extrasolar planets with orbits very close to the central star where
relativistic effects should have some signature.} {Our purpose is to
review how general relativity affects the orbital dynamics of the
planetary systems and to define a suitable relativistic correction
for Solar System orbital studies when only point masses are
considered.} {Using relativistic formulae  for the N body problem
suited for a planetary system given in the literature we present a
series of numerical orbital integrations designed to test the
relevance of the effects due to the general theory of relativity in
the case of our Solar System. Comparison between different
algorithms for accounting for the relativistic corrections are
performed.} {Relativistic effects generated by the Sun or by the
central star are the most relevant ones and produce evident
modifications in the secular dynamics of the inner Solar System. The
Kozai mechanism, for example, is modified due to the relativistic
effects on the argument of the perihelion. Relativistic effects
generated by planets instead are of very low relevance but
detectable  in numerical simulations.} \keywords{celestial mechanics
\and relativistic effects \and Solar System \and small bodies}
\end{abstract}

\section{Introduction}

General Relativity Theory (GRT), while widely used in some areas of
astrophysics, has not yet been fully exploited in the study of
planetary system dynamics, despite the fact that one of the most
famous relativistic effects, that of the precession of perihelia,
shows that Relativity can play a role, albeit somewhat secondary, in
the evolution of planetary systems.

The design of a detailed general relativistic reference frame for
use in the description of our Solar System was achieved in a
compelling though very technical way by Brumberg and Kopeikin and also by Damour et al. in a series of
publications \citep{kopeikin88,Brumberg89,Brumberg89b,Brumbergbook,damour,damour2,damour3}. In those papers, each
object in the system was equipped with a local reference frame (for
measuring internal degrees of freedom) which was fitted within a
global reference frame.  This work culminated in resolutions of the IAU 2000 General Assembly \citep{IAU}. The methods of Brumberg and Kopeikin, and Damour et al., allow
for the development of relativistic equations of motion for extended
objects, but for purposes of studying the general dynamics of
planetary systems, the limit of point (monopole) masses can be
expected to give good qualitative results.

  Also, there has been exhaustive studies of the validity of
GRT by means of the so called Parameterized Post Newtonian (PPN)
theories \citep{Will,Will93,klioner,Kopeikin2004} and comparison with
observations made in the Solar System (for a recent review, see \citet{Will2006}). Nevertheless, the effects measured for that
purpose are in general not related to any dynamical study of the
planetary system. There have also been a number of studies of Post
Newtonian effects, such as the Lens-Thirring effect in the motion of
satellites (see for example \citet{iorio} and also \citet{Ciufolini2007}), but again these were not
concerned with planetary dynamics studies.

Except for very precise simulations (for example \citet{Quinn},
\citet{varadi03}) relativistic effects are in general not taken into
account in the study of Solar System dynamics, and when they are,
only the effects due to the Sun are considered, using a
simplification of the above mentioned more complete PPN model found in \citet{Will}. This simplified approximation was also used
in the study of some extrasolar planetary systems. For example,
 \citet{Barnes} studied the stability of
the $\upsilon$-Andromedae planetary system using that approximation
and \citet{nagasawa} showed strong differences between classic and
relativistic evolutions of planet b of that system. A more general
study of the contribution of the relativistic terms to the secular
evolution of extrasolar systems was done by \citet{adams06}. In the case of extrasolar planets orbiting Pulsars, a specific study of relativistic dynamics is still pending, as stated in \citep{Gozdziewski05}.

In the case of minor bodies, as explained by \citet{saye94} GRT
 should be used when studying low perihelia orbits. They
also pointed out that inconsistences arise when using newtonian
dynamics with masses derived from relativistically derived
ephemerides.

Apart from the aforementioned studies, relativistic effects were included in the pioneering work at high precision ephemerides at the JPL, and by now they have a measurable impact on observations, with today's high precision astronomical techniques (see e.g. \citet{Pitjeva2005} for recent advances in the numerical integrations of planetary ephemerides).

It seems without doubt that future development of dynamical astronomy and experimental gravitation will fundamentally rely on the study of General Relativity effects. This is a common point of view expressed in resolutions of the IAU 2000 General Assembly \citep{IAU}

  The aim of this paper is to evaluate how GRT affects the
orbital dynamics of the Solar System, in the limit of point masses,
by means of a series of numerical tests. By extension we can extract
some conclusions for the extrasolar systems. Hopefully, the numerical studies here presented will help to elucidate in which circumstances it is necessary to account for relativistic effects.

\section{The Post-Newtonian Algorithm}\label{postnewt}

 We are interested in the first corrections to the classical
Newtonian equations of motion of a system of point masses, as
derived from the GRT. They can be found by means of an approximation
of this last theory, known as the Post Newtonian approximation. This
approximation can be accomplished by making an expansion of the GRT
equations in terms of $\mathbf{v}/c$, where the $\mathbf{v}$ are the
velocities, and $c$ is the speed of light. Equation of motion can
thus be found for point masses, which include the Newtonian term, as
well as a number of other terms which are suppressed by factors of
$1/c$. The PN equation of motion used in this work is given by
\citet{newhall83}:

\begin{eqnarray}
\mathbf{\ddot{r}}_i & = & \sum_{j \neq i} \frac{\mu_j (\mathbf{r}_j
- \mathbf{r}_i)}{r_{ij}^3} \bigg[1 - \frac{4}{c^2} \sum_{k \neq i}
\frac{\mu_k}{r_{ik}} - \frac{1}{c^2} \sum_{k \neq j}
\frac{\mu_k}{r_{jk}} +
\frac{\mathbf{v}_i^2}{c^2} + \nonumber \\
& & +2\frac{\mathbf{v}_j^2}{c^2} - \frac{4}{c^2} \mathbf{v}_i \cdot
\mathbf{v}_j - \frac{3}{2c^2} \left( \frac{(\mathbf{r}_i - \mathbf{r}_j) \cdot \mathbf{v}_j}{r_{ij}} \right)^2 + \frac{1}{2c^2} (\mathbf{r}_j - \mathbf{r}_i) \cdot \mathbf{\ddot{r}}_j\bigg] + \nonumber\\
& & + \frac{1}{c^2} \sum_{j \neq i} \frac{\mu_j}{r_{ij}^3}
\left[(\mathbf{r}_i - \mathbf{r}_j) \cdot (4 \mathbf{v}_i - 3
\mathbf{v}_j)\right] (\mathbf{v}_i - \mathbf{v}_j) + \frac{7}{2c^2}
\sum_{j \neq i} \frac{\mu_j \mathbf{\ddot{r}}_j}{r_{ij}}
\label{choclo}
\end{eqnarray}

  \noindent with $\mu_i$ standing for $Gm_i$, $\mathbf{v}_i$
is the barycentric velocity of $m_i$ and dots denoting usual
derivation with respect to an inertial time, making
$\mathbf{\ddot{r}}_j$ the acceleration of $m_j$. Here, ``velocity'', and ``acceleration'' must be understood in a coordinate sense. These quantities do not in fact correspond exactly to the velocity and acceleration as measured by any physical observer - a subtle difference not found in Newtonian Mechanics. The
summation is over all massive bodies including the Sun and the
origin is the barycenter of the system. This expression is known as
the Einstein-Infeld-Hoffman equation, and the first version of it is
found in \citet{EIH}. For a modern deduction, see \citet{Brumberg2007}.  The expression also coincides with the one given by \citet{damour} in the case of point monopole masses. A pedagogical derivation of this equation,
starting from GRT, is given in appendix
\ref{appendix}

As it stands, this algorithm would greatly slow down any numerical
calculation, and it is only useful in situations where the
relativistic contribution of \emph{every} object in the system is to
be taken into account. For the simplified case where only one object
(e.g. the Sun) contributes with relevant corrections, the sums in
(\ref{choclo}) are changed into a single term, and in the reference
frame of this central, massive object with mass $\mathcal{M}$, the
relativistic correction would read

\begin{equation}
\Delta \mathbf{\ddot{r}} = \frac{G\mathcal{M}}{r^3c^2} \left[
\left(\frac{4G\mathcal{M}}{r} - \mathbf{v}^2 \right) \mathbf{r} +
4(\mathbf{v} \cdot \mathbf{r })\mathbf{v} \right] \label{simple}
\end{equation}

which is the correction proposed by \citet{ander75} (in fact, the equation as written in \citet{ander75} depends on two PPN parameters
$\beta$ and $\gamma$, which are equal to unity in the GRT case) and used by most papers that consider relativistic
corrections \citep{Quinn,saye94,varadi03}. This algorithm, although
less exact than Eq. (\ref{choclo}), is computationally much more
affordable and, as we will show, allows for the most important
effects. Anyway, in certain cases the complete algorithm
(\ref{choclo}) may become necessary, when there are more than just
one relativistically relevant object in the system under
consideration. Sometimes, in very simplified analysis, the second term in
the right hand side of (\ref{simple}) is ignored but this can only
be justified when considering near circular orbits.

In going from equations (\ref{choclo}) to (\ref{simple}) one is not
taking into account possible relativistic effects due to the changes
in the velocity of the central object caused by other, large but not
necessarily relativistically relevant, objects. In the case of the
Solar System, the influence of Jupiter on the motion of the Sun
could in principle introduce such an effect, but a simple order of
magnitude analysis shows that this effect is highly suppressed and
can be consistently taken out.

\begin{figure}
\resizebox{\hsize}{!}{\includegraphics{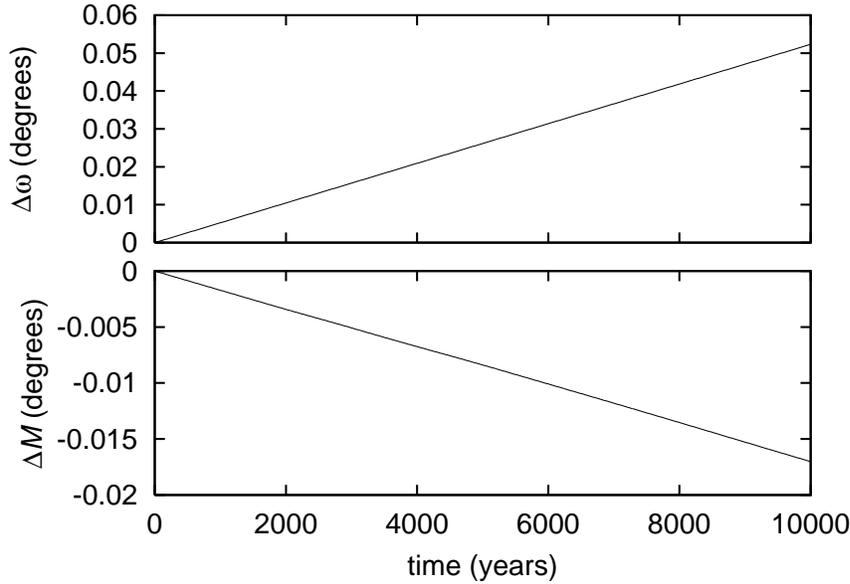}}
\caption{ A numerical integration considering only the Sun and a
massless particle with orbital elements similar to 2P/Encke
($a=2.22, e=0.85$) designed to show the relativistic effects on
$\omega$ and $M$ due to Eq. (\ref{simple}). Variations are $\Delta
\omega= \omega_{rel}-\omega_{cla}$ and $\Delta M= M_{rel}-M_{cla}$.
The short period effects are undetectable.
}\label{encke}
\end{figure}

It is possible to show that the only  secular contribution --that
means non short period-- from relativistic effect generated by the
central star affects only the argument of the perihelion $\omega$
and the mean anomaly $M$ which is related to an effect on the time
of perihelion passage as measured from a far observer at rest. The effect on $\omega$ due to the
relativistic effects of the Sun is given by $\Delta\omega =
0.0384/(a^{5/2}(1-e^{2}))$ arcseconds per year where $a$ is in AUs
\citep{Will,sitarski83,saye94}. For low eccentricity orbits the
variation in $M$ as measured by a far observer at rest is given approximately by $\Delta M \sim
-0.115/(a^{5/2}\sqrt{1-e^{2}})$ arcseconds per year and it is
possible to show that the short period effects are negligible
\citep{iorio}. Both secular effects should be considered, for
example, in the case of comets when obtaining their non
gravitational forces from the observed temporal evolution of the
orbital elements \citep{yeomans05}. These analytical predictions due
to pure relativistic effects can be easily verified by means of a
numerical integration of a massless particle orbiting the Sun under
the effects of Eq. (\ref{simple}). We show as an example a numerical
integration of a particle with orbital elements similar to 2P/Encke
at Fig. \ref{encke}. The short period effects are undetectable if
any at all, and the effects on $\omega$ are exactly as predicted.
The effects on mean anomaly $M$ are not exactly as predicted because
of the high value of $e$. The other orbital elements remain
unchanged. In the figure the time is as measured from a distant observer at rest.

Some sympletic numerical integrators use an artificial separable
Hamiltonian that can mimic the effects on
$\omega$ and $M$ that Eq. (\ref{simple}) generates \citep{satre}.
Using this artifact the relativistic effects can be included in the
perturbing Hamiltonian and sympletic form is maintained but
 this approximation, usually used in studies of our planetary system,
 is only valid for nearly constant and not very high eccentricity orbits.

We are interested in exploring how  relativistic effects generated
by Eqs. (\ref{choclo}) and (\ref{simple}) affect the dynamical
evolution of our planetary system and the dynamics of small bodies.
In order to analyze the relativistic effects in the Solar System we
implemented three different models: one using classic newtonian
accelerations (model C), another  which considers only the
relativistic effects generated by the Sun (R$_S$) and the last with
the expression that takes into account the relativistic effects
generated by all massive bodies (R$_{all}$).

\section{Effects in Our Planetary System}

Our planetary system is mainly composed by an inner set of
terrestrial low mass planets and an outer set of giant very massive
planets. As they evolve in different timescales we will analyze both
two groups separately. Our model for planetary system includes the
Sun and the eight mayor planets from Mercury to Neptune and we
consider the Earth-Moon system as unique body located at its
baricenter. This model is simpler than, for example, the one of
\citet{Quinn} so our results will not be more precise. Our aim is to
study the relevance of the relativistic corrections in order to have
an idea if they are or not important in modeling the orbital
dynamics of bodies evolving in the Solar System.

\subsection{The inner planetary system}

The inner planetary system is strongly perturbed by the giant
planets and it is known that the relativistic terms due to the Sun
have signatures in timescales of millions of years
\citep{laskar88,Quinn}. Are the relativistic effects generated by
the planets (model R$_{all}$) relevant or on the contrary are the
Sun's relativistic effects  (model R$_S$) enough to explain the
dynamics of the inner planetary system?

In order to investigate this point we have integrated all the
planetary system from Mercury to Neptune by 2 millon years. We
performed three different runs according to the three different
models: 1) model C using MERCURY integrator \citep{chambers} with
the Bulirsch--Stoer method including only classical newtonian
gravitation, 2) model R$_S$ using the same integrator with BS method
including an user defined acceleration given by Sun's relativistic
effects according to Eq. (\ref{simple}) and 3) model R$_{all}$ with
our own fully relativistic routine implemented using RA15
\citep{ever} that includes relativistic effects from all massive
bodies according to Eq. (\ref{choclo}). Runs with models C and R$_S$
were repeated for control using the classic and relativistic
versions of the integrator EVORB (\citet{fer02},
http://www.fisica.edu.uy/$\sim$gallardo/evorb.html).

\begin{figure}
\resizebox{\hsize}{!}{\includegraphics{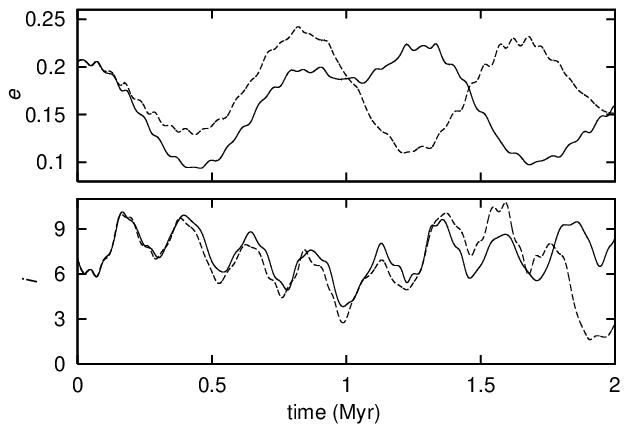}}
\caption{Mercury's evolution. Top: eccentricity; bottom: inclination.
Full line: classical model; dashed line: both relativistic models.}\label{mercury}
\end{figure}

\begin{figure}
\resizebox{\hsize}{!}{\includegraphics{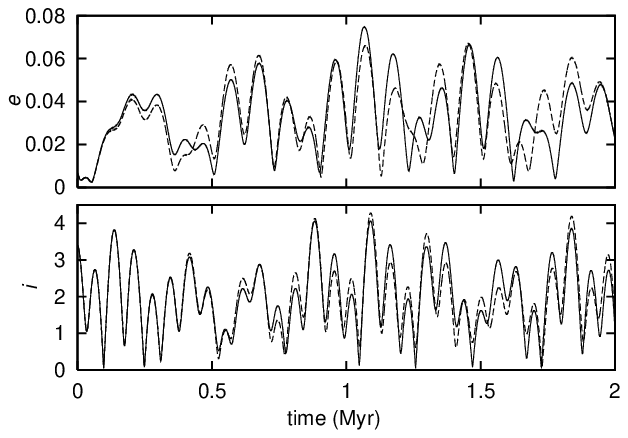}}
\caption{Venus's evolution. Top: eccentricity; bottom: inclination.
Full line: classical model; dashed line:  both relativistic models.
}\label{venus}
\end{figure}

\begin{figure}
\resizebox{\hsize}{!}{\includegraphics{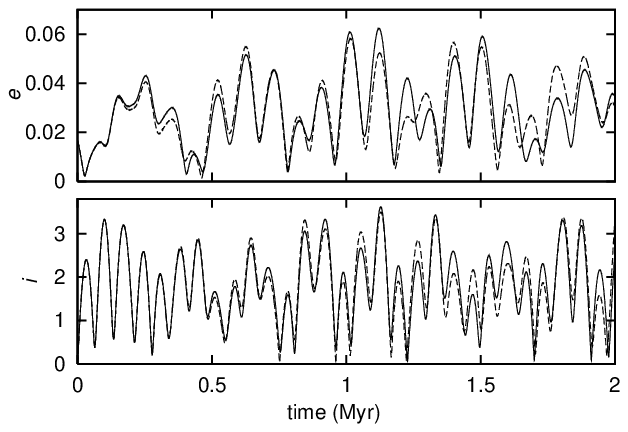}}
\caption{Earth's evolution. Top: eccentricity; bottom: inclination.
Full line: classical model; dashed line:  both relativistic models.}\label{earth}
\end{figure}

\begin{figure}
\resizebox{\hsize}{!}{\includegraphics{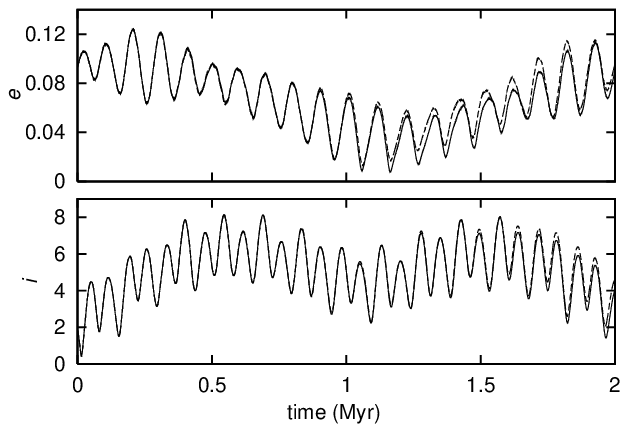}}\caption{Mars' evolution. Top: eccentricity; bottom: inclination. Full
line: classical model; dashed line:  both relativistic models.}
 \label{mars}
\end{figure}

Figures \ref{mercury}-\ref{mars} show the evolutions of the  planets
from Mercury to Mars. It is possible to see that the main results
from \citet{laskar88} are recovered. Notable differences between
classic and relativistic models are evident in the case of Mercury
(Fig. \ref{mercury}). In the case of Venus (Fig. \ref{venus}) and
the Earth (Fig. \ref{earth}), the relativistic integrations present
also some variations with respect to the classical one in the
evolution of eccentricities and inclinations, in addition to a
precession of the perihelion. For Mars, effects are of much lower
importance and start to be appreciable at the end of the
integration, also mainly in the evolution of eccentricity and
inclination.

For the considered time intervals the integration results from both
relativistic models R$_{S}$ and R$_{all}$ were basically
undistinguishable. We want to stress that both relativistic models
R$_{S}$ and R$_{all}$ were implemented in very different numerical
integrators with very different algorithms (BS and RA15
respectively), so the coincidence of the results is a prove of their
precision and a confirmation that the differences detected with
model C are real and not a numerical artifact.

We tested the relevance of the relativistic effects by running an
integration that included Earth plus the outer planetary system
(that is without the less massive terrestrial planets) for a time
span of 2 millon years. By doing this, we eliminate the forced modes
generated by the other terrestrial planets and we isolate the
relativistic effects from the inner Solar System chaos
\citep{laskar96}. The resulting evolution of Earth's eccentricity
and specially the inclination are much less affected by the
relativistic corrections. Then, the discrepancies between classic
and relativistic evolutions found in the real Earth in Fig.
\ref{earth} are mainly due to the complexity that characterize the
dynamics of the complete inner planetary system, making the small
relativistic effects on the inner planets to interact in an complex
way.

\subsection{The outer planetary system}

We have shown that relativistic effects generated by the Sun
 are relevant but the ones generated by the planets are negligible in the inner planetary system. Can we conclude
the same for the outer more massive planetary system?
 We are interested in relativistic effects due to the
massive outer planetary system and to compare with the relativistic
effects due exclusively to the Sun, that means model R$_{all}$
versus R$_S$. In this experiment we integrated the outer planetary
system from Jupiter to Neptune by 401 millon years. We performed
three different runs according to the three different models defined
earlier.

\begin{figure}
\resizebox{\hsize}{!}{\includegraphics{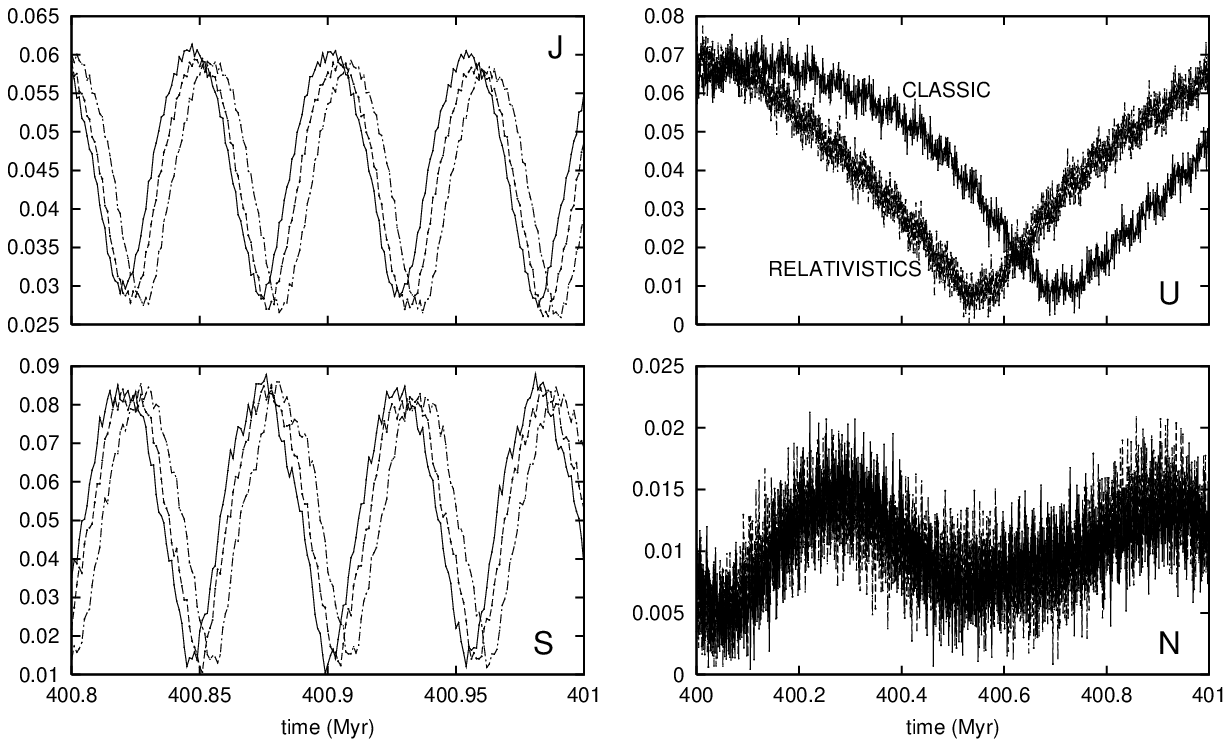}}
\caption{Eccentricity evolution of the outer planetary system. Full
line: classical model C; dashed line: relativistic model R$_S$ ;
dashed dotted line: relativistic model R$_{all}$.}\label{jsune}
\end{figure}

\begin{figure}
\resizebox{\hsize}{!}{\includegraphics{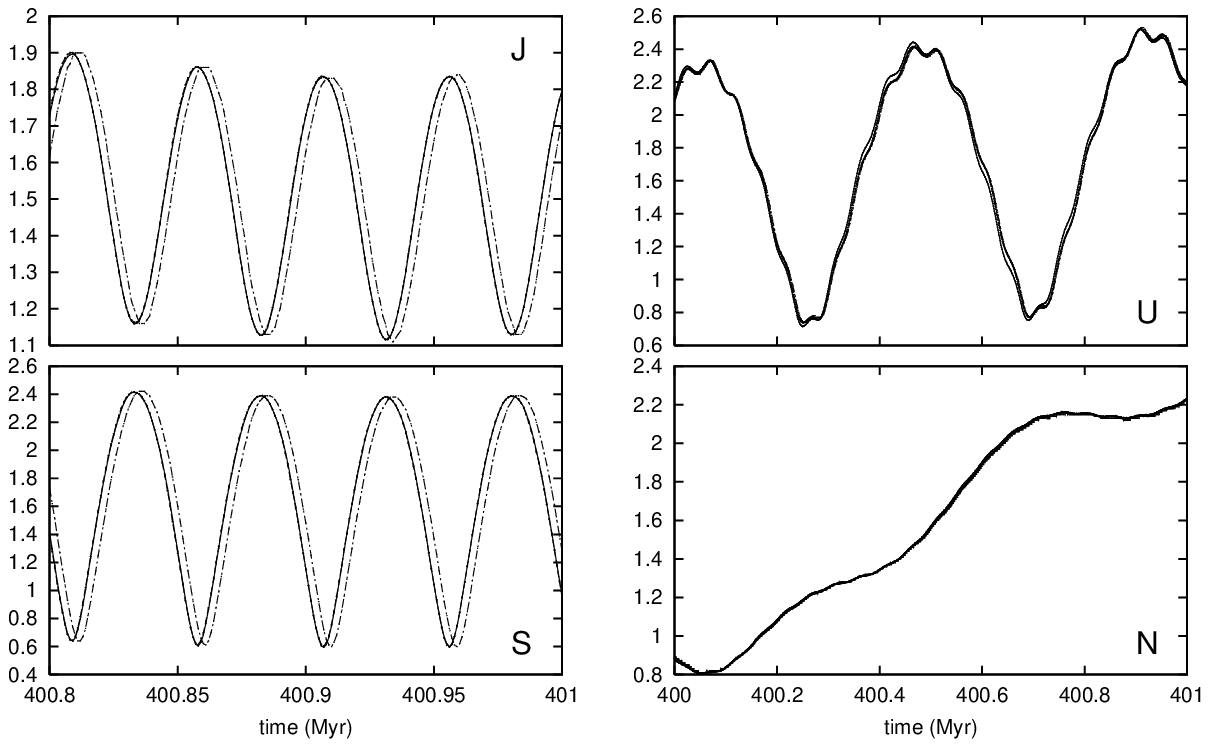}}
\caption{Inclination evolution of the outer planetary system. Full
line: classical model C; dashed line: relativistic model R$_S$ ;
dashed dotted line: relativistic model R$_{all}$.
}\label{jsuni}
\end{figure}

In Figs. \ref{jsune} and \ref{jsuni} we show the last years of the
evolution of the eccentricity and inclination respectively for the
four giant planets. There is a small shift in $e$ and a very small
shift in $i$ between the different models which should affect very
slightly the fundamental frequencies \citep{laskar88} but the global
behavior is not substantially changed.

Results for the more massive and near Jupiter and Saturn are
somehow different than results for the less massive and remote
planets Uranus and Neptune. Analyzing the eccentricity showed at Fig.
\ref{jsune} we can see that model R$_{all}$ departs from R$_{S}$ and
both from the classic model C. In Uranus both relativistic models
are almost coincident and they depart from C whereas for Neptune
differences are something small. Analyzing the inclinations for Jupiter and
Saturn in Fig. \ref{jsuni} we can see that model C and R$_S$
coincide and they depart from R$_{all}$, so mutual relativistic effects between Jupiter and Saturn seem to be
more important than the relativistic effects of the Sun on Jupiter and Saturn. No differences in
inclination are appreciated in Uranus and Neptune. We can interpret
this as follows: the two most massive and near planets Jupiter and
Saturn generate some additional perturbations in model R$_{all}$ but
these effects are not enough in the more distant and less massive
planets Uranus and Neptune. Effects in the inclination are of lower
magnitude than in the eccentricity and only in the more massive and
near planets R$_{all}$ show up.

It is known that quasi resonant angles like $2\lambda_J -
5\lambda_S$ or $2\lambda_N - \lambda_U$ where $\lambda$ is the mean
longitude can generate non negligible perturbations on minor bodies
\citep{femiro98,mife01,marza}. A modification of the time evolution
of the quasi resonant angles due to relativistic effects could
introduce dynamical effects in the time evolution of these minor
bodies. Analyzing our integrations we conclude that the circulation
periods of both critical angles remain practically unchanged by the
relativistic terms, so that there do not exist relevant relativistic
effects on these angles.

\section{Effects in Solar System's Low $(a,q)$ Orbits}\label{lowq}

\subsection{Secular dynamics of fictitious particles}

According to the expression for the relativistic effect
$\Delta\omega$ we can infer that the low $(a,q)$ orbits should be
the most sensible to the relativistic effects, and according to the
results of section 3.1, if we exclude planet crossing orbits, we can ignore the relativistic effects
generated by the planets in the case of low $(a,q)$ orbits.
So, we have integrated by $10^5$ years
the system composed by the Sun, the four giant planets and a
population of test particles with low perihelion ($0.05<q< 0.5$ AU)
and low semimajor axis ($0.2<a<2$ AU) with and without the
relativistic terms due to the Sun, that is, models R$_S$ and C
respectively. We also performed a run with the model R$_S$ including the
terms corresponding to the solar quadrupole moment of the Sun
$J_2=2\times 10^{-7}$ \citep{pir03} obtaining a completely
negligible effect, so these effects can be ignored in comparison
with the relativistic effects due to the Sun.

\begin{figure}
\resizebox{\hsize}{!}{\includegraphics{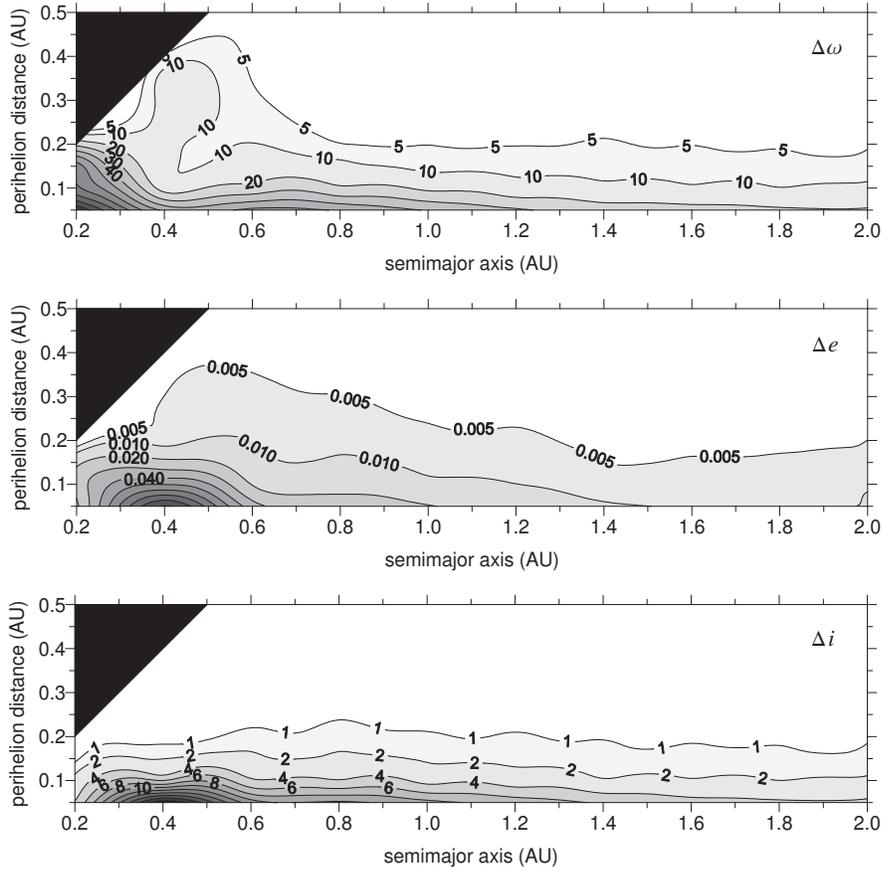}}\caption{Maximum differences between classic and relativistic secular
dynamics generated by the giant planets as function of the initial
$(a,q)$ values in a numerical integration for $10^{5}$ years. All
test particles have initial $i=20^{\circ}, \Omega=0^{\circ},
\omega=60^{\circ}$. The extreme changes observed in $(e,i)$ at
$a\sim 0.4$ are due to a strong Kozai mechanism (see Fig. 9). These
differences increase when considering the secular effects of the
entire planetary system.}
\label{superficies}
\end{figure}

We considered the time evolution of the elements $e,i,\omega$ and
computed the differences between both runs, classic and
relativistic. The maximum differences $\Delta\omega$, $\Delta e$ and
$\Delta i$ found during the integration are showed at Fig.
\ref{superficies} as a function of the initial semimajor axis and
perihelion distance. That figure was constructed with 100 test
particles all with initial $i=20^{\circ}, \Omega=0^{\circ},
\omega=60^{\circ}$, so it is not a map valid for any set of orbital
elements but give us an idea of how relativity modify the secular
dynamics in that region.

This experiment designed excluding the terrestrial planets
 avoids the chaotic dynamics typical of objects with
close encounters with them and the differences between both
runs give us and indication of the differences generated by
the relativistic terms in the secular, more regular, dynamics.

Besides some secular resonances present in this region
\citep{michel,mifro} the secular evolution that dominate this region
is the Kozai mechanism
 which produces strong variations in $(e,i)$ linked
to the evolution of $\omega$ maintaining a constant value of
$H=\sqrt{1-e^{2}}\cos i$, where the inclination is measured with
respect to the invariable plane of the Solar System. The
relativistic terms affect $\omega$ modifying the Kozai mechanism and
in consequence the time evolution of $(e,i)$  \citep{Morbidelli}.

\begin{figure}
\resizebox{\hsize}{!}{\includegraphics{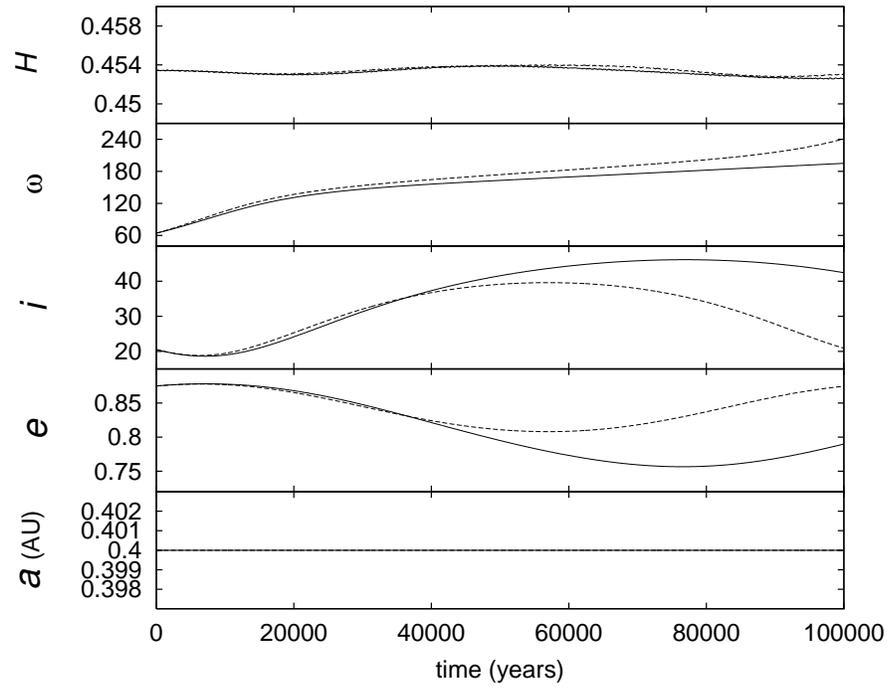}}\caption{Time evolution of a fictitious particle inside the secular
Kozai mechanism generated by the giant planets. Full line: classical
model C; dashed line: relativistic model R$_S$. In spite of
considerable variations in $e$ and $i$ the parameter
$H=\sqrt{1-e^2}\cos i$ remains nearly constant in both models. Relativistic
terms diminishes the amplitude of the oscillations in $(e,i)$.}
 \label{particula14}
\end{figure}

At Fig. \ref{particula14} we show an example of how relativity
modifies the Kozai mechanism. It is a typical secular evolution with
constant $a=0.4$ AU in both models within all the time integration
period. Strong differences between the other orbital elements appear
conserving $H$ nearly constant.

In the real Solar System these effects can be increased several
times due to the secular forced modes generated by the terrestrial
planets but in the case of orbits having close encounters with the planets
the chaotic dynamics will overcome any improvement to the model introduced
by the relativistic corrections.

\subsection{The known relativistic asteroid population}

After the successfully explanation of  the precession of perihelion
of Mercury by Einstein's GRT, an
unavoidable  step was to study the evolution of the first very low
perihelion asteroid known, 1566 Icarus \citep{shapiro68}, where
relativistic effects were expected and  observed in its dynamical
evolution. Icarus, being maybe the most studied ``relativistic''
asteroid, is by no means the only one to deserve such treatment.
Mercury is the Solar System's body with greatest relativistic
$\Delta\omega$ (0.4304 $"/\mathrm{yr}$) but there are at present
around half hundred asteroids with $\Delta\omega$ greater than
Icarus ($\Delta\omega=0.1006$ $"/\mathrm{yr}$ ). The objects with
the greatest values for $\Delta\omega$ are shown at Table
\ref{topten} which was constructed from the Near Earth Asteroids
database of the Minor Planet Center and  can be considered as an
update of Table 1 of \citet{saye94}. These asteroids are the ones
with the most important relativistic effects on $\omega$ but not
necessarily the ones with the most diverging evolutions from classic
model C. Divergence depends also on the degree of chaos of their
dynamics.

\begin{figure}
\resizebox{\hsize}{!}{\includegraphics{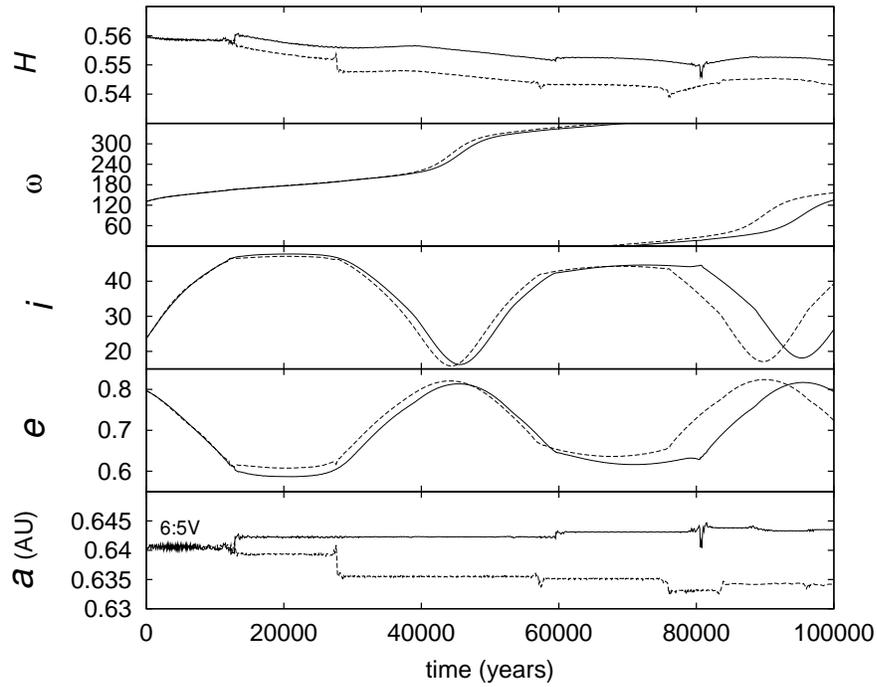}}
\caption{Asteroid 2004 XY60. At the beginning the asteroid
is in the 6:5 resonance with Venus. After 13000 years the strength
of the resonance drops and both the classic and relativistic
particles are captured in different high order mean motion
resonances with Earth.}\label{xy60}
\end{figure}

We have integrated the full planetary system plus all these
asteroids and also some others for a time span of $10^5$ years using
both the classical and relativistic models C and R$_S$. All
asteroids from the table exhibit chaotic evolutions with several
close encounters with terrestrial planets that generate  quick
departures between classic and relativistic trajectories. Resonance
sticking \citep{duncan97,lykawka06} is also present with
 switches  between resonances that can be easily identified following
 \citet{gall}. Asteroid 2004 XY60 is a good example (Fig.
\ref{xy60}) of an evolution dominated by close encounters plus
resonance sticking. It starts inside resonance 6:5 with Venus and
after 13000 years the resonance's strength drops and high order
resonances with Earth start to dominate following different
dynamical trajectories the classic and the relativistic models.

\begin{figure}
\resizebox{\hsize}{!}{\includegraphics{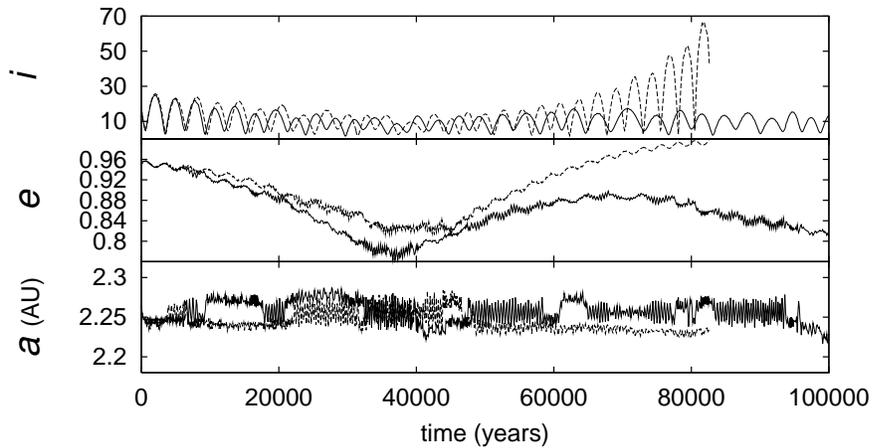}}
\caption{Asteroid 2000 LK. An example where relativistic effects plus
chaotic dynamics generate very different evolutions. Full line:
classical model C; dashed line: relativistic model R$_S$. The
relativistic simulation ends up colliding with the Sun. In this case
the parameter $H$ is not conserved.}\label{LK}
\end{figure}

\begin{figure}
\resizebox{\hsize}{!}{\includegraphics{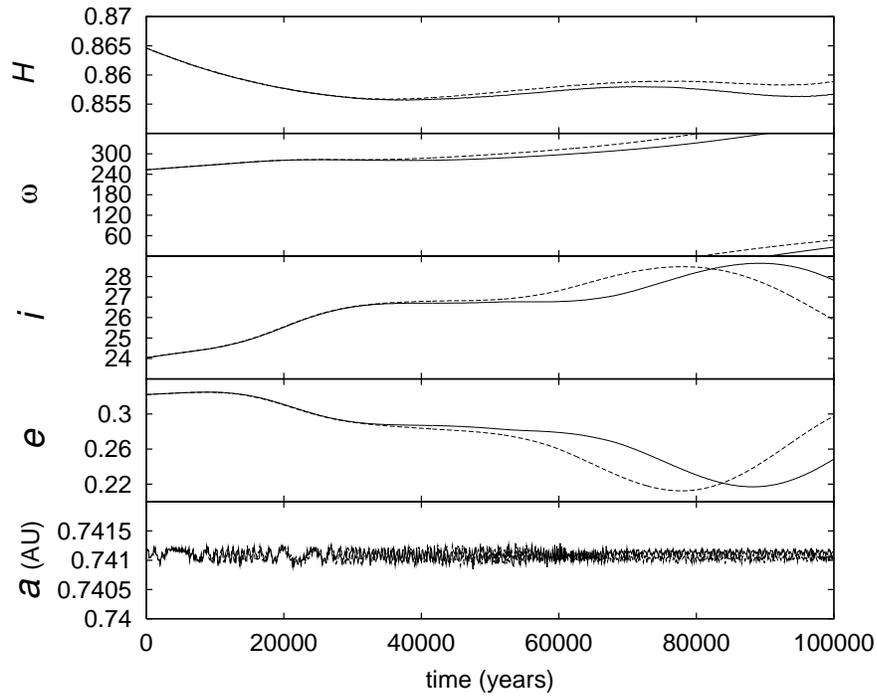}}
\caption{Asteroid 2003 CP20. An example of evolution without
encounters with the planets. Full line: classical model C; dashed
line: relativistic model R$_S$. The asteroid is captured in
resonance 27:28 with Venus and follows the effects of the Kozai
mechanism.}\label{cp20}
\end{figure}

\begin{figure}
\resizebox{\hsize}{!}{\includegraphics{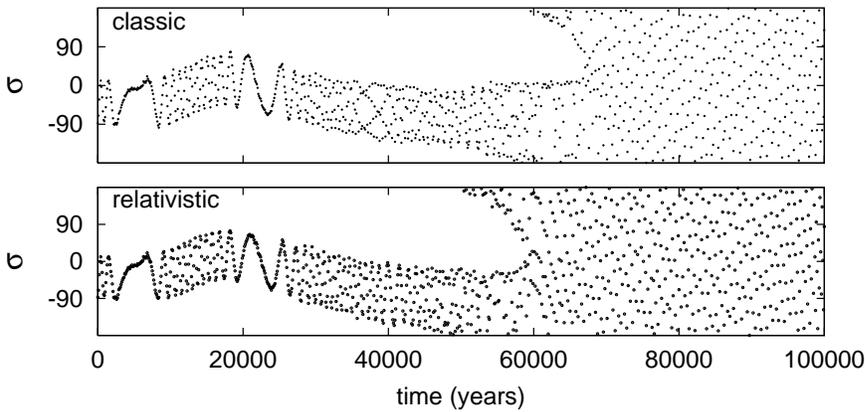}}\caption{Asteroid 2003 CP20. The critical angle of the resonant
motion is $\sigma=-27\lambda_{Ven}+28\lambda-\varpi$ according to
the classic and relativistic models.}
\label{res}
\end{figure}

As an extreme case we can mention
 2000 LK, not included in Table \ref{topten}, in which the relativistic particle hits the Sun after 83000 years,
and the classical one survive during the whole integration (Fig.
\ref{LK}). A more interesting case is 2003 CP20 (Fig. \ref{cp20})
neither included in the table, which did not have any close
encounter with planets in any of the runs but it is strongly
affected by exterior resonance 27:28 with Venus as we have found
following \citet{gall}. The critical angle
$\sigma=-27\lambda_{Ven}+28\lambda-\varpi$ is shown at Fig.
\ref{res} for the two models C and R$_S$. In this case the Kozai
mechanism is also present generating important variations in $e,i$
in analogy to other situations observed in the transneptunian region
\citep{gomes}.

We have also studied some real asteroids and fictitious particles in
mean motion resonances with Jupiter obtaining  very different
evolutions between models C and R$_S$. This is an expected result
because the chaotic nature of the resonant dynamics makes that small
variation in the models to generate exponential divergences.
 For completeness we have studied the dynamical evolution
of fictitious particles in the transneptunian region but no
relativistic effects were found there.

\section{Extrasolar Systems}

The relevance of relativistic effects in extrasolar systems were
already shown by, for example, \citet{mardling} and
\citet{nagasawa}. In particular \citet{nagasawa} showed that
relativistic terms have dramatically affected the evolution of
planet b in $\upsilon$ Andromedae. \citet{adams06} have extended the
secular interaction theory including GR terms and they found that
relativistic corrections are relevant for systems with close planets
($\sim$4 days orbits). Then, it is not necessary to insist here on
the importance of including relativistic corrections specially when
semimajor axes are of the order of $10^{-1}$ AU or less, but we can
analyze the relevance of the use of model R$_{all}$ instead of
R$_{S}$.

With this in mind we integrated a fictitious planetary system with
low eccentricity and close orbits of massive planets (see Table
\ref{sistema}) using the three models.
We avoided experiments with very close orbits in order to avoid
chaotic dynamics.
The global dynamics obtained
by the three models are very similar and regular due to the small
eccentricity of the orbits. Differences are mainly in the
frequencies of the secular evolution. We have fourier analyzed one
of the fundamental frequencies of the system obtaining
$f=1.63677\times 10^{-3} \mathrm{yr}^{-1}$ for model C,
 $f=1.63682\times
10^{-3} \mathrm{yr}^{-1}$ for model R$_{S}$ and $f=1.63681\times
10^{-3} \mathrm{yr}^{-1}$ for model R$_{all}$. We have checked that
the differences were  not a numerical artifact repeating the
simulation R$_{all}$ with lower precision parameter for RA15 and
verifying that the new frequency was exactly the same. Then,
differences between models  R$_{S}$ and  R$_{all}$ seem to be real
and not due to numerical artifacts but, at least in this example,
negligibly small. We repeated the experiment with the same planetary system but
in this case with smaller semimajor axes ($0.2$ and $0.8$ AU) and an analogue behavior was observed regarding
the differences between fundamental frequencies in the three models.

However, relativistic effects due to the planets (model  R$_{all}$)
could be important for planetary systems in mean motion resonances
or under the mechanism of perihelion alignment or anti-alignment
 evolving very near the separatrix in the phase space
because a small difference in the evolution of $\varpi$ could
generate a very different behavior in the  phase space with an
associated very different evolution in $e$.

The problem of the relevance of model R$_{all}$ deserves a more
detailed study. In principle we can conclude that in planetary
systems, even with low $e$, it is worthwhile to consider
relativistic corrections at least with model R$_{S}$ and a test with
model R$_{all}$ will be convenient.

\section{Conclusions}

 The secular relativistic effects generated by the Sun are relevant
for the argument of the perihelion and mean anomaly (related to the
time of perihelion passage) whereas the short period effects are
vanishingly small. In our planetary system relativistic effects
generated by the Sun are appreciable specially in the inner Solar
System,  but those generated by the planets are very small and are
only detectable in the long time scales in the giant planets,
specially Jupiter and Saturn.

Due to the secular dynamics the departure between classic and
relativistic trajectories is enhanced in the inner Solar System. In
particular the Kozai mechanism is substantially modified driving
notorious differences in the evolution of the eccentricity and
inclination.

Nowadays there exist a noticeable population of relativistic minor
bodies in the Solar System. Around a half hundred asteroids at
present exhibit greater relativistic effects than 1566 Icarus. The
relativistic corrections due to the Sun  are enough for the correct
modeling of the dynamics of low perihelion or low semimajor axis
bodies. Minor bodies evolving having close approaches to the planets
or being in  resonances are the most affected by the relativistic
corrections not because of the magnitude of the relativistic corrections but due to the chaotic nature of their dynamics.
Relativistic corrections should also be applied to comets but in
this case in order to be consistent a suitable model for the non
gravitational forces is necessary.

In general, in extrasolar planets the relativistic corrections due
to the star are enough and, in some cases,  are indispensable as in
the case of systems evolving near a separatrix of trajectories in
the phase space. In some particular cases the relativistic effects
generated by the planets could have some importance, but this point
deserves future research.

In the case of exoplanets orbiting around pulsars, a more detailed study needs to be performed. Such a study could be done following the same lines as those presented in this paper, and it constitutes the obvious direction for future research. As for the case of binary pulsar systems, its study is more complex and the algorithms used in this paper would need to be improved.

Some years ago \citet{sitarski83} concluded ``...it seems that in all
the modern investigations it is the very time to replace Newtonian
equations of motions by those resulting from general relativity
theory''. Following the results here presented, it seems that relativistic corrections are not unavoidable for all
dynamical studies, but they are necessary for the precise dynamical
modeling of bodies evolving in the inner Solar System and for the
dynamical modeling of close extrasolar planets.

\begin{acknowledgements}
 We thank Pablo Mora for its invaluable help in the early
stages of this work.
      This work was done in the framework of Proyecto CSIC "Dinamica Secular de Sistemas Planetarios y Cuerpos
      Menores".
\end{acknowledgements}

\appendix

\section{Appendix: A Closer Look at the Post-Newtonian Algorithm}\label{appendix}

General Relativity is a theory written in a language quite unlike
the one of newtonian gravitation. In GR, space--time is equipped
with a \emph{metric} $g_{\mu \nu}$, which measures distances and
angles at every point. There does not exist any gravitational
\emph{force} in the theory, and the gravitational interactions are
given by the form of the metric itself. Free particles in the theory
follow \emph{geodesics}, that is, trajectories which minimize
distances as given by the metric.

Nevertheless, as any new theory, GR must, and of course does,
recover the classical theory as a limit of weak gravitational fields
and low velocities. This is what is called the \emph{newtonian
limit} of the theory, and in it the classical field $\phi$ is
recovered as a term in the temporal part of the metric $g_{\mu \nu}$
of space--time (\citet{Wein,Gravitation})

\begin{displaymath}
g_{00} = -(1 + 2\phi)
\end{displaymath}

In our case, we take this limit and improve it using an expansion in
the typical velocities of the particles of the system $\mathbf{v}^2
\sim GM/r$ , following \citet{Wein}. For velocities which are small
compared with the speed of light $c$, that means $GM/r c^2 \ll 1$,
this approximation is sufficient to describe the dynamics of point
masses in the Solar System. Following the usual convention in GR, we
work in units where $c=1$. We can then write

\begin{eqnarray}\label{desarrollo1}
g_{0 0} & = & -1 + g_{0 0}^{(2)} + g_{0 0}^{(4)} + \ldots \nonumber \\
g_{ij} & = & \delta_{ij} + g_{ij}^{(2)} + g_{ij}^{(4)} + \ldots \nonumber \\
g_{i0} & = & g_{i0}^{(3)} + g_{i0}^{(5)} + \ldots \nonumber
\end{eqnarray}
where numbers in parenthesis denote the order in $\mathbf{v}/c$ of
the terms, and the Latin indexes represent the spatial components.
Notice that, in order to preserve properties of symmetry under time
reversal at this order of the expansion (which is a valid symmetry, due to the fact that gravitational radiation effects are of a higher order in the expansion \citep{Wein,Gravitation}, only terms with a given parity are allowed
in the expansion.

Similar expansions can be produced for the inverse metric tensor
$g^{\mu \nu}$ and for the Ricci tensor of curvature $R_{\alpha
\beta}$. In GR, curvature of space--time is given by second
derivatives of the metric tensor, and measures how much geodesics
are different to straight lines. The relation between these
expansions can be readily deduced, giving expressions in the form of

\begin{displaymath}
R_{ij}^{(2)} = - \frac{1}{2} \frac{\partial^2 g_{0
0}^{(2)}}{\partial x^i \partial x^j} + \frac{1}{2} \frac{\partial^2
g_{kk}^{(2)}}{\partial x^i \partial x^j} - \frac{1}{2}
\frac{\partial^2 g_{ik}^{(2)}}{\partial x^k \partial x^j}
-\frac{1}{2} \frac{\partial^2 g_{kj}^{(2)}}{\partial x^k
\partial x^i} + \nabla^2 g_{ij}^{(2)}
\end{displaymath}

\noindent for the spatial components, making use of Einstein's summation convention.

These expressions can be simplified by the use of some arbitrary
\emph{gauge conditions}, which are permitted due to the invariance
properties of GR theory under arbitrary coordinate transformations.
This kind of freedom is analogous to those that exist in
electromagnetic theory. In our case we followed \citet{Wein} in
choosing the so-called \emph{Harmonic Coordinate Conditions}, which
force the metric to satisfy

\begin{displaymath}
\Gamma^{\lambda} \equiv g^{\mu \nu} \Gamma^{\lambda}_{\mu \nu} = 0
\end{displaymath}

\noindent where $\Gamma^{\lambda}_{\mu \nu}$ are the Christoffel
symbols, related to the first derivatives of the metric, and also to
the parallel transport of vectors along a curve in space--time. The use of
these conditions impose restrictions to our expansion terms, e.g.

\begin{displaymath}
0 = \frac{1}{2} \frac{\partial g^{(2)}_{00}}{\partial t} -
\frac{\partial g^{(3)}_{i0}}{\partial x^i} + \frac{1}{2}
\frac{\partial g^{(2)}_{ii}}{\partial t}
\end{displaymath}

\noindent these restriction help to simplify our equations.

Then we can proceed to make use of \emph{Einstein's equation}, which
relates in each point the curvature of space--time to the presence
of matter and energy:

\begin{displaymath}
R_{\mu \nu} = -8 \pi G (T_{\mu \nu} - \frac{1}{2} g_{\mu \nu}
T^{\lambda}_{\lambda}) \qquad \textrm{with} \qquad T_\lambda^\lambda
= g^{\mu \lambda} T_{\mu \lambda}
\end{displaymath}

\noindent where $T_{\mu \nu}$ is the \emph{energy--momentum} tensor,
including the density of energy in space--time. In the case of
point particles, this density will reduce to a summation over
Dirac's $\delta$ functions at the position of each particle. This
tensor can also be expanded in terms of $\mathbf{v}/c$, and when we
substitute the expansions in Einstein's equation we reach equations
of the form

\begin{eqnarray}
\nabla^2 g^{(2)}_{00} & = & -8 \pi G T^{(0) 00} \nonumber\\
\nabla^2 g^{(4)}_{00} & = & \frac{\partial^2g^{(2)}_{00}}{\partial
t^2} + g^{(2)}_{ij} \frac{\partial^2 g^{(2)}_{00}}{\partial x^i
\partial x^j} - \left( \frac{\partial g^{(2)}_{00}}{\partial x^i}
\right)
\left( \frac{\partial g^{(2)}_{00}}{\partial x^i} \right) \nonumber \\
 & & -8\pi G \left[T^{(2) 00} - 2g^{(2)}_{00} T^{(0) 00} + T^{(2) ii}\right] \nonumber\\
\nabla^2 g^{(3)}_{i0} & = & 16 \pi G T^{(1) i0} \nonumber\\
\nabla^2 g^{(2)}_{ij} & = & -8 \pi G \delta_{ij} T^{(0) 00}
\nonumber
\end{eqnarray}

And we find, as was to be expected

\begin{displaymath}
g^{(2)}_{00} = -2 \phi = 2G \int \frac{T^{(0) 00}(\mathbf{r'},
t)}{|\mathbf{r} - \mathbf{r'}|} d^3r'
\end{displaymath}

\noindent where $T^{(0)00}$ is just the mass density of the system,
and $\phi$ the classical potential.

We can now define some new potentials, which will play the role of
corrections to the classical theory.

\begin{displaymath}
g^{(3)}_{i0} = \zeta_i
\end{displaymath}

where

\begin{displaymath}
\zeta_i(\mathbf{r},t) = -4G \int T^{(1)i0} \frac{(\mathbf{r'},
t)}{|\mathbf{r} - \mathbf{r'}|} d^3r'
\end{displaymath}

\begin{displaymath}
g^{(4)}_{00} = -2\phi^2 - 2\psi
\end{displaymath}

where

\begin{eqnarray}
\psi(\mathbf{r}, t) & = & - \int \frac{1}{|\mathbf{r} -
\mathbf{r'}|} \bigg[ \frac{1}{4\pi} \frac{\partial^2
\phi(\mathbf{r'},t)}{\partial t^2} + GT^{(2) 00}(\mathbf{r'},t) +
GT^{(2) ii}(\mathbf{r'},t) \bigg] d^3r' \nonumber
\end{eqnarray}

It is possible now to write a relativistic equation of motion in
terms of these potentials. As we have already stated, in GR test
particles move following geodesics, and the equation for a geodesic
is given, in terms of a generic curve length parameter $\tau$ by

\begin{displaymath}
\frac{d^2x^\mu}{d\tau^2} + \Gamma^{\mu}_{\nu\lambda}
\frac{dx^\nu}{d\tau} \frac{dx^\lambda}{d\tau} = 0
\end{displaymath}

\noindent where the Christoffel symbols, which are needed for making
covariant derivatives in a curved space--time, are also written in terms of
the metric, and of its expansion.

Splitting up of temporal and spatial indexes leads to the expression

\begin{eqnarray}
\frac{d^2x^i}{dt^2} & = & - \Gamma^{i}_{0 0} - 2\Gamma^{i}_{0j}
\frac{dx^j}{dt} - \Gamma^{i}_{jk} \frac{dx^j}{dt} \frac{dx^k}{dt}  +
\left[ \Gamma^{0}_{0 0} + 2\Gamma^{0}_{0j} \frac{dx^j}{dt} +
\Gamma^{0}_{jk} \frac{dx^j}{dt} \frac{dx^k}{dt}\right]
\frac{dx^i}{dt} \nonumber
\end{eqnarray}

\noindent finally, after recovering the factors of $c$ for a more
intuitive expression, substitution of the three potentials give

\begin{eqnarray}
\frac{d \mathbf{v}}{dt} & = & - \nabla (\phi + 2\frac{\phi^2}{c^2} +
\psi) - \frac{1}{c} \frac{\partial \mathbf{\zeta}}{\partial t} +
\frac{\mathbf{v}}{c} \times ( \mathbf{\nabla} \times \mathbf{\zeta})+ \nonumber \\
& & +  \frac{3}{c^2} \mathbf{v} \frac{\partial \phi}{\partial t} +
\frac{4}{c^2} \mathbf{v} (\mathbf{v \cdot \nabla})\phi -
\frac{\mathbf{v}^2}{c^2} \mathbf{\nabla}\phi \nonumber
\end{eqnarray}

Given this equation of motion, it becomes possible to understand the
physical meaning of the three potentials, $\phi$, $\psi$ and
$\zeta$. The case of $\phi$ is easiest, as it is nothing but the
classical gravitational potential from Newton's theory, thus
recovering the classical limit when $\mathbf{v} \ll c$. The vector
potential $\zeta$ is quite similar to the vector potential
$\mathbf{A}$ in electrodynamics. Just as the magnetic field can be
seen as a (special) relativistic effect of the electric field, so
$\zeta$ appears as a relativistic correction to the classical field,
which exerts a velocity--dependent force normal to the distance
vector and to the relative velocity of two massive objects. This
field is sometimes known in the literature as the
\emph{gravitomagnetic} field, for its obvious analogies with the
magnetic field in electrodynamics.

The $\psi$ scalar field contains corrections to $\phi$ due to many
relativistic effects. It can be shown that the time derivative term
in the $\psi$ defining integral accounts for the delay of the
gravitational signal, which travels at the speed of
light, and which moves following a geodesic instead of a straight
line. The other two terms are corrections due to the equivalence
between mass and energy, by means of which the kinetic and potential
energies of the particles modify the classical field.

The equation of motion contains the classical equation of Newton as
its first term, and many small corrections to it. The equation in
itself can prove to be very insightful for a theorist, but it is not
fit yet for use in a numerical algorithm. We would like to find a
more straightforward algorithm, in which the acceleration of each
point particle would be given explicitly by the positions and
velocities of every particle in the system.

To accomplish this we must use the relativistic expression for the
momentum--energy tensor of a system of particles, in order to find
$\phi$, $\psi$ and $\zeta$ as functions of positions and velocities.
In the case of $\psi$, the delay--of--signal term can prove very
hard to simplify. Some methods and approximations from
electrodynamics can be used, as the delay--of--signal term is almost
identical to the Lenard--Jones electric potential (e.g
\citet{Griffiths}). After some algebra it is possible to arrive to
the formula given by \citet{newhall83}, which is more explicit --
and numerically useful -- albeit somewhat less physically
meaningful:

\begin{eqnarray}
\mathbf{\ddot{r}}_i & = & \sum_{j \neq i} \frac{\mu_j (\mathbf{r}_j
- \mathbf{r}_i)}{r_{ij}^3} \bigg[1 - \frac{4}{c^2} \sum_{k \neq i}
\frac{\mu_k}{r_{ik}} - \frac{1}{c^2} \sum_{k \neq j}
\frac{\mu_k}{r_{jk}} +
\frac{\mathbf{v}_i^2}{c^2} + \nonumber \\
& & +2\frac{\mathbf{v}_j^2}{c^2} - \frac{4}{c^2} \mathbf{v}_i \cdot
\mathbf{v}_j - \frac{3}{2c^2} \left( \frac{(\mathbf{r}_i - \mathbf{r}_j) \cdot \mathbf{v}_j}{r_{ij}} \right)^2 + \frac{1}{2c^2} (\mathbf{r}_j - \mathbf{r}_i) \cdot \mathbf{\ddot{r}}_j\bigg] + \nonumber\\
& & + \frac{1}{c^2} \sum_{j \neq i} \frac{\mu_j}{r_{ij}^3}
\left[(\mathbf{r}_i - \mathbf{r}_j) \cdot (4 \mathbf{v}_i - 3
\mathbf{v}_j)\right] (\mathbf{v}_i - \mathbf{v}_j) + \frac{7}{2c^2}
\sum_{j \neq i} \frac{\mu_j \mathbf{\ddot{r}}_j}{r_{ij}} \nonumber
\end{eqnarray}

\noindent which is the same as eq. (\ref{choclo}) in the main text.

\newpage

\begin{table}
\caption{Solar System's bodies most affected by relativistic effects
generated by the Sun. } \label{topten} \centering
\begin{tabular}{cccccc}
\hline \hline
Designation & $a$ (AU) & $e$ & $i (^{\circ})$ & $\Delta\omega$ ("/yr) \\
\hline
Mercury   & 0.387     & 0.206 & 7.0 &   0.430  \\
2004 XY60   & 0.640     & 0.797 & 23.7 &   0.321  \\
2000 BD19   & 0.876     & 0.895 & 25.6 &   0.268  \\
2006 CJ   &  0.676  & 0.755 & 10.2 &   0.238  \\
(66391) 1999 KW4    &  0.642  & 0.688 & 38.8 &   0.221  \\
1995 CR   &  0.907  & 0.869 &  4.0 &   0.200  \\
1999 MN   & 0.674   & 0.665 &  2.0 &   0.185  \\
2006 KZ39   &  0.616    & 0.525 &  9.3 &   0.178  \\
2001 TD45   & 0.797     & 0.777 & 25.4 &   0.171  \\
2004 JG6   & 0.635  & 0.531 & 18.9 &   0.166  \\
1998 SO   & 0.731   & 0.699 & 30.3 &   0.164  \\
(85953) 1999 FK21    & 0.739   & 0.703 & 12.5 &   0.162  \\
2004 XZ130   & 0.618    & 0.455 &  2.9 &   0.161  \\
2006 JF42   & 0.672     & 0.582 &  5.9 &   0.157  \\
2005 WS3   & 0.672  & 0.576 & 23.0 &   0.155  \\
2004 UL   & 1.266   & 0.927 & 23.7 &   0.151  \\
\hline
\end{tabular}
\end{table}

\begin{table}
\caption{A fictitious planetary system adopted to evaluate the
effects of models C, R$_S$ and R$_{all}$. The mass of the central
star was taken equal to $M_{\odot}$. } \label{sistema} \centering
\begin{tabular}{ccccc}
\hline \hline
Object & $a$(AU) & $e$ & $i$($^{\circ}$) & $m/m_{Jup}$ \\
\hline
Planet b   & 0.6     & 0.04 & 1.04 &   1.0  \\
Planet c   & 1.0     & 0.02 & 1.45 &   1.0  \\
\hline
\end{tabular}

\end{table}

\newpage

\end{document}